\documentclass[aps,prd,nofootinbib,floatfix,twocolumn]{revtex4-2}
\usepackage{amsmath,amsfonts,amssymb,color,graphicx,bm,hyperref,mathrsfs}
\usepackage{xcolor}
\usepackage{mathptmx}
\usepackage{physics}
\usepackage{framed}
\usepackage{orcidlink}

\begin{document}
\title{\textbf{Entropy of matter on the Carroll geometry}}
\author{Saurav Samanta\orcidlink{0009-0009-9498-5739}$^{1,}$}
\email{srvsmnt@gmail.com}
\affiliation{$^1$Department of Physics, Bajkul Milani Mahavidyalaya, Kismat Bajkul, Purba Medinipur 721655, India.}
\author{Bibhas Ranjan Majhi\orcidlink{0000-0001-8621-1324}$^{2,}$}
\email{bibhas.majhi@iitg.ac.in}
\affiliation{$^2$Department of Physics, Indian Institute of Technology Guwahati, Guwahati 781039, Assam, India.}

\begin{abstract}
Two prescriptions for the construction of Carroll geometries, the expansion of geometric variables near horizon and the expansion of metric with zero limit of the expansion parameter $c$ (speed of light in vacuum), are known to complement each other. The entropy of an ideal gas, confined in a box and kept very close to the horizon, depends on the transverse area of the container. We show this by using the Carroll geometry constructed through the expansion of the metric and then taking the zero limit of the expansion parameter $c$. Therefore, the present analysis assures the complementing nature of two ways of finding the Carroll geometry from the thermodynamical point of view.

\end{abstract}
\maketitle	

The Theory of relativity reduces to Newtonian theory when the speed of light in free space tends to infinity. Recently, the other extreme limit, which is $c\to0$, attracted a lot of attention. The latter is called the Carrollian limit, as the Poincare symmetry group maps to the Carroll group \cite{Levy-Leblond:1965dsc,SenGupta:1966qer,Bacry:1968zf,Henneaux:1979vn,deBoer:2017ing}. In this limit the Minkowski light cone closes as $(dx/dt) = c\to0$ along the light ray and the theory which belongs to this ultra-local regime is consequently called Carrollian theory. Contrary to Galilean transformations, the space is absolute and time is relative in Carroll transformations. Just to mention, there can be other types of Carroll particles with zero energy but they are not spatially fixed \cite{deBoer:2021jej}.

The features of Carrollian geometry can also be mimicked by the geometry of a null surface on a generic spacetime (see, e.g. \cite{Penna:2018gfx,Donnay:2019jiz,Redondo-Yuste:2022czg,Ecker:2023uwm,Ciambelli:2018xat,Ciambelli:2018ojf,Bhattacharya:2024fbz}). One such important null-surface is the horizon. The reason is as follows. Approaching towards horizon also leads to closing of the light cone. For example, consider the Rindler metric    
\begin{eqnarray}
ds^2=e^{2a\xi/c^2}(-c^2d\eta^2+d\xi^2)+dy^2+dz^2~,
\label{B1}
\end{eqnarray}
where Minkowski coordinates $(t,x,y,z)$ and Rindler coordinates $(\eta,\xi,y,z)$ are related through the following relations:
\begin{eqnarray}
&&t = \frac{c}{a}e^{\frac{a\xi}{c^2}}\sinh\Big(\frac{a\eta}{c}\Big)~;
\nonumber
\\
&&x = \frac{c^2}{a}e^{\frac{a\xi}{c^2}}\cosh\Big(\frac{a\eta}{c}\Big)~.
\label{B2}
\end{eqnarray}
In the above, the parameter $a$ is the acceleration of the frame. Here, the Killing horizon is given by $\xi\to -\infty$, in which case the horizon is defined by $x=\pm ct$ (as the trajectory is given by $x^2 - c^2t^2 = (c^4/a^2) e^{2a\xi/c^2}$). In this case, one can easily verify that $\tanh(a\eta/c) = ct/x = \pm 1$ and hence at the horizon we have $(a\eta/c)\to\pm\infty$. This last limit can be achieved by imposing $c\to 0$, leading to $dx/dt\to0$ --- closing of the light cone. 

Another example will be of interest to us -- the static spherically symmetric spacetime in Schwarzschild coordinates $(t,r_s,\theta,\phi)$:
\begin{equation}
ds^2 = - f(r_s)c^2dt^2 + \frac{dr_s^2}{f(r_s)}+r_s^2d\Omega^2~,
\label{B3}
\end{equation}
where $d\Omega^2 = d\theta^2 + \sin^2\theta d\phi^2$. It admits a horizon at $r_s=r_H$ for which $f(r_H)=0$. Radial null rays are given by $ds^2=0$ with constant $\theta$ and $\phi$. So, these null rays, which form the light cone, are determined through $(dr_s/dt) = \pm cf(r_s)$. Therefore, $r_s\to r_H$ leads to $(dr_s/dt)\to 0$, implying the closing of the light cone. 

Both of the above examples illustrate the common fact that spacetime with horizon can play a pivotal role in accomplishing the Carrollian physics. In particular, the geometry near the horizon successfully mimics the Carrollian structure. This interpretation has been adopted in various situations and is found to be fruitful (e.g., Carroll hydrodynamics \cite{Ciambelli:2018xat,Ciambelli:2018ojf,Petkou:2022bmz,Freidel:2022bai,Redondo-Yuste:2022czg,Bhattacharya:2024fbz}, Carrollian motion of particle \cite{Gray:2022svz,Ciambelli:2023tzb}). These studies led to two points of view to achieve the Carrollian version of a geometry -- (a) construction of geometry near a null surface, like the horizon, with null surface approaching limit and (b) expansion of the background metric with $c^2=0$, known as Carrollian geometry. The first approach is very old and it has been associated with Carrollian regime in recent times. This has been adopted in various cases, like the fluid-gravity correspondence of gravity \cite{Ciambelli:2018xat,Ciambelli:2018ojf,Petkou:2022bmz,Freidel:2022bai,Bhattacharya:2024fbz}, the accelerated world-sheet of a closed string to tensionless string \cite{Bagchi:2020ats,Bagchi:2021ban,Karan:2024jgn}, etc. The other formalism is also equally fruitful in achieving a Carroll extension of various theories. In this case, the theory is constructed on the Carroll expanded metric and then terms up to leading order in $c^2$ are kept(see, e.g., \cite{Bagchi:2023cfp,Banerjee:2025bkg} and references therein). Both formalisms appeared to complement each other as in the $c\to0$ limit the Carroll expanded metric, used in the second formulation, is mimicked by the near horizon expansion adopted in the first formalism. 

In this regard, our main interest in this article is to investigate the thermodynamics of matter near a horizon. In the literature various studies in this direction have been done \cite{Kolekar:2010py,Bhattacharya:2017bpl,Kothawala:2011fm,Sourtzinou:2022xhd,Samanta:2025wrn,Sourtzinou:2025hyd} and among them, the property of entropy of matter, when the system is placed very near the horizon, has a very specific feature. The entropy turns out to depend on the area of the container of matter, rather than its volume \cite{Kolekar:2010py,Bhattacharya:2017bpl,Sourtzinou:2022xhd,Samanta:2025wrn}. The observation has been tested for the Rindler and event horizons \cite{Kolekar:2010py}, as well as for cosmological horizons \cite{Bhattacharya:2017bpl}. This signifies the robustness of the distinct behavior of the entropy of matter in very strong gravity. As discussed earlier, the physics near the horizon is equivalent to taking the Carrollian limit $c\to0$ and therefore we categorize this observation as viewpoint (a). Naturally, we expect that the above-mentioned behavior of matter entropy near the horizon, as the viewpoint (b) suggests, can also be simulated by considering a direct application of Carroll limit $c\to0$. In search of a bridge between these viewpoints related to the features of thermodynamic matter entropy, here we investigate the entropy of a massless ideal gas having $N$ number of molecules confined in a box. In one case, the gas box is kept in a Rindler frame and the calculation is done in the Carroll limit $c\to0$. In the other case, the investigation is performed by considering the box on a Carroll expanded version of the Schwarzschild metric. In both situations, we directly apply the Carroll limit (i.e. $c\to 0$).

We observe that the dependence of the matter entropy on the cross-section of the box emerges quite naturally in both situations. The basic difference from the earlier analysis \cite{Kolekar:2010py,Bhattacharya:2017bpl} is in the application of the relevant limits. In the present study, we use the limit $c\to0$ instead of the near horizon limit. The similarity in the nature of entropy in these two ways has significant implications. First, it further strengthens the bridge between two viewpoints of the Carroll description. Second, as a consequence, for the first time we get the idea that such behavior of the thermodynamics of matter near the horizon is due to the underlying Carroll geometry. Third, the degrees of freedom which are responsible for the entropy effectively become Carrolian in structure and therefore the Carroll symmetries of the thermodynamical system may have a fundamental role to illuminate such area-dependent behavior. The fourth is the most ambitious one. The emergence of area-dependent nature of matter entropy may imply that the horizon might have an inherent Carroll structure. Since the entropy of a black hole itself is a function of horizon area, it might be due to some yet to be identified ``Carroll degrees of freedom''. Finally, the consistency of the two viewpoints illuminates the thermodynamic properties of Carroll matter.

Let us first briefly describe the basic formalism in the canonical ensemble to calculate the entropy of a thermodynamic system on a spacetime. It is elaborately described in \cite{Kolekar:2010py}. The partition function of the system at the inverse temperature $\beta$ is given by
\begin{equation}
Z=\int dE g(E)e^{-\beta E}~,
\label{B4}
\end{equation}
where the density of states $g(E)$ at energy $E$ is related to the phase-space volume $P(E)$ as $g(E)= dP(E)/dE$. For a static spacetime, equipped with a timelike Killing vector, the phase-space volume of a thermodynamic system consisting of massless particle turns out to be (see Eq. (5) of \cite{Kolekar:2010py} with $D=3$) 
\begin{eqnarray}
P(E)=\frac{4\pi}{3}E^3\int d^3x\frac{\sqrt{\gamma}}{(-g_{00})^{3/2}}~.
\label{B5}
\end{eqnarray}
In the above, $\gamma$ is the determinant of the spatial part of the background  metric and $g_{00}$ is the time-time component of the metric. The entropy is given by
\begin{equation}
S = \ln Z + \beta \bar{E}~, \,\,\,\,\ \bar{E} = -\frac{\partial\ln Z}{\partial\beta}~.
\label{B6}    
\end{equation}

In the following, we use the above formalism to find the entropy of a box of ideal gas consisting of $N$ number of massless non-interacting molecules. Two specific backgrounds will be considered -- one is the Rindler and the other is the Carroll expanded form of the Schwarzschild metric.

{\it Rindler metric: --}
We take the metric of the form (\ref{B1}). In this case, we have $\gamma=e^{2a\xi/c^2}$ and $g_{00}=-e^{2a\xi/c^2}$. Substituting these in (\ref{B5}) and integrating over the region of the container, one finds
\begin{eqnarray}
P(E)&=&\frac{4\pi}{3}E^3A_{\perp}\int_{\xi_a}^{\xi_b}d\xi e^{-\frac{2a\xi}{c^2}}
\nonumber
\\
&=& \frac{4\pi E^3}{3}\frac{A_{\perp}}{2}\frac{c^2}{a}e^{-\frac{2a\xi_a}{c^2}}\left[1-e^{-\frac{2aL}{c^2}}\right]~.
\label{B7}
\end{eqnarray}
In the above, $A_{\perp} = \int dydz$ is the transverse area of the box and $\xi_a$ and $\xi_b$ (with $\xi_a<\xi_b$) are the longitudinal positions of the 
two ends of it. We denote the longitudinal distance as $L=\xi_b-\xi_a$.
Now, the proper length distance ($L_{\textrm{prop}}$) of the $\xi_a$ surface of the box from the horizon is
\begin{eqnarray}
L_{\textrm{prop}}=\int_{-\infty}^{\xi_a}e^{\frac{a\xi}{c^2}}d\xi=\frac{c^2}{a}e^{\frac{a\xi_a}{c^2}}~.
\label{B8}
\end{eqnarray}
So in terms of this, $P(E)$ takes the following form:
\begin{eqnarray}
P(E)=\frac{4\pi E^3}{3}\frac{A_{\perp}}{2}\frac{c^6}{a^3L_{\textrm{prop}}^2} \left[1-e^{-\frac{2aL}{c^2}}\right]~.
\label{B9}
\end{eqnarray}

The carroll limit is given by $\frac{a}{c^2}\rightarrow \infty$ and so $1-e^{-\frac{2aL}{c^2}}\approx 1$. In this case, the phase-space volume is given by
\begin{eqnarray}
P(E)|_{\text{Carroll}}=\frac{2\pi E^3}{3}\frac{c^6}{a^3L_{\textrm{prop}}^2}A_{\perp}~.
\label{B10}
\end{eqnarray}
This leads to the following expression of entropy: 
\begin{equation}
S|_{\text{Carroll}} = N\Big[\ln\Big(\frac{4\pi c^6 A_\perp}{Na^3L^2_{\text{prop}}\beta^3}\Big)+4\Big]~.
\label{B11}    
\end{equation}
In the above, we have included the Gibbs' factor $N!$ due to the indistinguishability of gas molecules and also used the Stirling's approximation ($\ln N! \simeq N\ln N - N$) for large $N$ . Note that the entropy now depends on the transverse area of the box. 

However, in the Newtonian limit $\frac{a}{c^2}\rightarrow 0$, we have $1-e^{-\frac{2aL}{c^2}}\approx \frac{2aL}{c^2}$. Therefore, the phase-space volume depends on the volume of the container: 
\begin{eqnarray}
P(E)|_{\text{Newtonian}}=\frac{4\pi c^4 E^3}{3a^2 L^2_{\text{prop}}}V~,
\label{B12}
\end{eqnarray}
where $V= A_{\perp} L$ is the volume of the box. Therefore, as expected, the entropy in the Newtonian limit depends on the volume of the container.

{\it Carroll-Schwarzschild metric: --}
Now we consider the box in the Schwarzschild metric of the form (\ref{B3}) with $f(r_s) = 1-(r_H/r_s)$. The radial positions of the two ends of the box are $r_{s(a)}$ and $r_{s(b)}$ with $r_{s(a)}<r_{s(b)}$. The Carroll expanded form of the Schwarzschild metric is then obtained by substituting the transformation
\begin{eqnarray}
r_s=r_H+\frac{\epsilon r^2}{r_H}~ 
\label{R}
\end{eqnarray}
and then keeping the terms up to $\mathcal{O}(\epsilon)$ with identification $\epsilon \sim c^2$. With this, the Carroll-Schwarzschild metric turns out to be
\begin{eqnarray}
dS^2=r_H^2d\Omega^2+\epsilon\left[-\frac{r^2}{r_H^2}dt^2+4dr^2+2r^2d\Omega^2\right]+\mathcal O(\epsilon^2).
\label{B13}
\end{eqnarray}
The Carroll limit in this context is given by $\epsilon\rightarrow 0$ (for details, see \cite{Bagchi:2023cfp}).

Substituting $\sqrt{\gamma}\approx 2r_H^2\sqrt{\epsilon}~\sin\theta$ and $g_{00}=-\frac{r^2}{r_H^2}\epsilon$ in (\ref{B5}) one finds
\begin{eqnarray}
P(E)&=&\frac{8\pi E^3r_H^3}{3\epsilon}(\Omega r_H^2)\int_{r_a}^{r_b}\frac{dr}{r^3}
\nonumber
\\
&=&\frac{4\pi E^3r_H^3}{3\epsilon}(\Omega r_H^2)\frac{r_b^2-r_a^2}{(r_ar_b)^2}~.
\label{B14}
\end{eqnarray}
In the above $\Omega$ is the solid angle subtended by the transverse face of the box. $r_a$ and $r_b$ are determined through the relations
\begin{eqnarray}
&& r_{s(a)}=r_H+\frac{\epsilon r_a^2}{r_H}~;
\label{Ra}
\\
&& r_{s(b)}=r_H+\frac{\epsilon r_b^2}{r_H}=r_{s(a)}+L~,
\label{Rb}
\end{eqnarray}
where $L$ is the radial length of the box.
In terms of $r_a$ and $L$, $P(E)$ is expressed as
\begin{eqnarray}
P(E)=\frac{4\pi E^3r_H^3}{3}(\Omega r_H^2)\frac{L r_H}{r_a^2(\epsilon L r_H+\epsilon^2r_a^2)}~.
\label{B15}
\end{eqnarray}

Now in the Carroll limit $\epsilon\rightarrow 0$, $r_H^2=r_{s(a)}^2+\mathcal{O}(\epsilon)$ and therefore the leading order term of $P(E)$ is given by
\begin{eqnarray}
P(E)=\frac{4\pi E^3r_H^3}{3\epsilon r_a^2}A_{\perp}~
\label{B16}
\end{eqnarray}
where $A_\perp = \Omega r^2_{s(a)}$ is the transverse area of the $a$-face of the box, intercepted by the solid angle $\Omega$. 
The proper length distance of the $a$-face from the horizon is 
\begin{eqnarray}
L_{\textrm{prop}}=\int_0^{r_a}dr\sqrt{4\epsilon}=\sqrt{4\epsilon}~r_a~.
\label{B17}
\end{eqnarray}
In terms of this, the phase-space volume is given by
\begin{eqnarray}
P(E)=\frac{16\pi E^3r_H^3}{3L_{\textrm{prop}}^2}A_{\perp}~.
\label{B18}
\end{eqnarray}
So, the calculated entropy from this again depends on the transverse area of the container of the gas.

A generalization to any static spherically symmetric metric of the form (\ref{B3}) can be easily done. For completeness, we summarize this below.

{\it Carroll-static spherically symmetric metric: --}
Using the coordinate transformation 
\begin{equation}
r_s = r_H + 2\epsilon \kappa r^2~,
\label{B19}
\end{equation}
on (\ref{B3}) and keeping up to $\mathcal{O}(\epsilon)$ we obtain the Carroll expanded form for any spherically symmetric metric as
\begin{eqnarray}
dS^2\simeq r_H^2d\Omega^2 + 4\epsilon\left[-\kappa^2r^2 c^2dt^2+dr^2+\kappa r_Hr^2\Omega^2\right]~.
\label{B20}
\end{eqnarray}
Here we define $\kappa = f'(r_H)/2$.
Following identical steps, one finds the following.
\begin{eqnarray}
P(E)=\frac{\pi E^3}{6\kappa^3}(\Omega r_H^2)\frac{Lr_H}{r_a^2(\epsilon L r_H+\epsilon^2r_a^2)}~.
\label{B21}
\end{eqnarray}
Finally in the limit $\epsilon\rightarrow 0$, keeping the leading order term, we find
\begin{eqnarray}
P(E)=\frac{\pi E^3}{6\epsilon \kappa^3 r_a^2}A_{\perp}~.
\label{B22}
\end{eqnarray}
In terms of the proper radial distance from the horizon of the $a$-face $L_{\text{prop}} = 2\sqrt{\epsilon}~r_a$, the above becomes
\begin{eqnarray}
P(E) = \frac{2\pi E^3}{3\kappa^3}\frac{A_{\perp}}{L^2_{\text{prop}}}~.
\label{B23}
\end{eqnarray}
The Area dependence of phase-space further indicates that the entropy is a function of the transverse area of the box.

An interesting point to note is that if $L_{\text{prop}}$ is identified with Planck length $L_p$ the expressions for the phase-space, given by (\ref{B10}) and (\ref{B23}), have one-to-one resemblance with the corresponding expressions obtained by near horizon analysis (see Eqs. (16) and (31) of \cite{Kolekar:2010py}). This is consistent as one can check that $L_{\text{prop}}$ becomes very small in the limit $c\sim\sqrt{\epsilon}\to 0$. Also it may be mentioned that $\kappa$ in (\ref{B23}) has the dimension of inverse-length. The same is also true for $a/c^2$, appeared in (\ref{B10}). In fact if one compares the Rindler transformations (\ref{B2}) with near horizon Kruskal transformations for a static-spherically symmetric metric (given by Eq. (8.16) in Section $8.3$ of \cite{book:PadmanabhanGrav}), these will be equivalent provided the surface-gravity $\kappa$ is mapped to acceleration  parameter through $\kappa\to a/c^2$. Therefore for comparison between expressions (\ref{B10}) and (\ref{B23}) one needs to keep this analogy in mind. Once this is considered, both are in equal footings.  

The thermodynamic entropy of a box of gas (consisting of massless, non-interacting $N$-number of molecules) has been investigated on a spacetime which inherently captures a horizon structure. Specifically, few simplified spacetimes have been considered -- one is the Rindler spacetime and another one is the Carroll expanded form of Schwarzschild spacetime. Extension to the Carroll expanded form of an arbitrary static, spherically symmetric black hole has been done as well. We found that in the Carroll limit (i.e. $c\to0$) the entropy depends on the cross-sectional area of the box, rather than its volume. This observation nicely complements the previous analysis \cite{Kolekar:2010py,Bhattacharya:2017bpl,Samanta:2025wrn} done near the horizon and thereby re-establishes the bridge between two ways of constructing the Carroll geometry. 

The previous argument on the dependence of entropy on the cross-sectional area is as follows. The strong gravity near the horizon contracts the longitudinal length of the container so that the degrees of freedom reduce to two, i.e. the gas effectively feels it is living on two-dimensional space. The present analysis shows that the same effect can be mimicked by performing the analysis on the Carroll geometry constructed using the $c\to0$ limit. In this regard, the construction of this geometry can also be done by considering the near null surface region. In particular, one defines the geometry around the horizon through timelike foliation and the evolution of the  geometrical degrees of freedom, lying on the foliating hypersurface, resulting in the Carroll structure. This idea is being evolved through recent progress in the Carroll fluid-gravity correspondence \cite{Ciambelli:2018xat,Ciambelli:2018ojf,Petkou:2022bmz,Freidel:2022bai,Bhattacharya:2024fbz}. This hypersurface is $(2+1)$-dimensional for a $(3+1)$-dimensional manifold. Here, as we obtain the area dependence of the entropy by using Carroll geometry, we feel that the present observation can further shed light on the contraction of one space dimension due to the strength of gravity. Therefore, it can lead to a deeper understanding of the role of the horizon in inducting the distinct behaviour of the thermodynamics of matter. However, to arrive at a concrete answer, further investigation is required, which we leave for the future.

\bibliographystyle{apsrev}

\bibliography{bibtexfile}

\end{document}